\RequirePackage{fix-cm}
\documentclass[twocolumn, epjc3]{svjour3}  
\usepackage{amsmath,amssymb}
\usepackage{color}

\smartqed  
\RequirePackage{graphicx}
\journalname{Eur. Phys. J. C}
\begin{document}

\title{Universal relation involving fundamental modes in two-fluid dark matter admixed neutron stars
}


\author{Hajime Sotani\thanksref{e1,addr1,addr2,addr3}
        \and
        Ankit Kumar\thanksref{e2,addr1} 
}

\thankstext{e1}{e-mail: sotani@yukawa.kyoto-u.ac.jp}
\thankstext{e2}{e-mail: ankitlatiyan25@gmail.com}


\institute{Department of Mathematics and Physics, Kochi University, Kochi, 780-8520, Japan \label{addr1}
           \and
           RIKEN Center for Interdisciplinary Theoretical and Mathematical Sciences (iTHEMS), RIKEN, Wako 351-0198, Japan \label{addr2}
           \and
           Theoretical Astrophysics, IAAT, University of T\"{u}bingen, 72076 T\"{u}bingen, Germany\label{addr3}
}

\date{Received: date / Accepted: date}

\maketitle

\begin{abstract}
We systematically investigate the fundamental oscillation frequencies of dark matter admixed neutron stars, focusing on models with self-interacting fermionic dark matter that couples to normal matter solely through gravity. The analysis is carried out within a two-fluid formalism under the relativistic Cowling approximation, where the perturbation equations follow from the linearized energy-momentum conservation laws of both components. We find that the mass-scaled fundamental frequencies of the nuclear (dark) fluid in dark core (halo) configurations exhibit a remarkably tight correlation with the total stellar compactness. This universality persists across the dark matter parameter space explored in this study and is largely insensitive to the choice of nuclear equation of state. In contrast, we also find the breakdown of such universality with the tidal deformability, i.e, the same frequencies show substantial deviations from universality when expressed in terms of the tidal deformability. These contrasting behaviors highlight possible observational imprints of dark matter in neutron star interiors.
\end{abstract}

\section{Introduction}
\label{sec:I}

Neutron stars, the compact remnants of core-collapse supernovae, provide unique environments for studying physics under extreme conditions. Their central densities exceed nuclear saturation, and their gravitational and magnetic fields surpass those observed anywhere in the solar system~\cite{ST83}. Due to the uncertain equation of state (EOS) at supranuclear densities, key structural properties of neutron stars—such as mass and radius—remain poorly constrained. Observational data from neutron stars and their associated phenomena thus play a critical role in constraining the EOS at high densities.

Discoveries of neutron stars with masses around $2M_\odot$ have already ruled out the soft EOSs that predict maximum masses below the observed values~\cite{D10,A13,C20,F21}. Future detections of even more massive neutron stars would further tighten these constraints. Because neutron stars are intrinsically relativistic objects, photons emitted from their surfaces experience strong gravitational bending, producing characteristic relativistic effects. This relativistic light bending significantly shapes the observed pulsar light curves, which are primarily sensitive to the stellar compactness, defined as $M/R$, where $M$ and $R$ denote the stellar mass and radius, respectively~\cite{PFC83,LL95,PG03,PO14,SM18,Sotani20a}. In practice, the mass and radius of neutron stars have been constrained through Neutron star Interior Composition Explorer (NICER) observations of PSR J0030+0451~\cite{Riley19,Miller19} and PSR J0740+6620~\cite{Riley21,Miller21}. We note that the constraint on PSR J0030+0451 has been updated recently~\cite{Vinciguerra24}.
In parallel, gravitational-wave detections have emerged as a powerful tool for probing neutron star properties, complementing traditional electromagnetic observations. In particular, the direct detection of gravitational waves from the binary neutron star merger GW170817 has placed stringent limits on the tidal deformability, thereby constraining the radius of a $1.4\ M_{\odot}$ neutron star to $R_{1.4}\lesssim 13.6$ km~\cite{GW170817,Annala18}. While these astronomical observations mainly constrain the EOS at supranuclear densities, terrestrial experiments provide complementary constraints at lower densities~\cite{SNN22,SO22,SN23}. Notably, by combining constraints on nuclear saturation properties with quasi-periodic oscillations (QPOs) observed in the magnetar flare GRB 200415A, the mass and radius of the corresponding neutron star have also been estimated~\cite{SKS23}.

Oscillation frequencies of neutron stars also play an important role in probing their internal properties. These objects exhibit discrete oscillation modes, the so-called eigenmodes, whose frequencies are highly sensitive to the star’s internal structure. When new physical ingredients are introduced into the system, additional oscillation modes may be excited. Therefore, by identifying the oscillation frequencies with specific modes, as an inverse problem, one can extract information about the underlying physics. This technique is known as asteroseismology, analogous to the well-established seismology on Earth and helioseismology for the Sun. For instance, the QPOs observed from the magnetar giant flares are thought to originate from neutron star oscillations. By associating such observed QPO frequencies with specific neutron star oscillation modes---particularly crustal torsional oscillation modes---the neutron star mass, radius, and crust EOS have been constrained, e.g.,~\cite{GNHL2011,SNIO2012,SIO2016,SKS23,Sotani24a}. In a similar way, the detection of the gravitational waves from neutron stars offers a powerful means to constrain their properties, such as mass, radius, EOS, and rotational characteristics, e.g.,~\cite{AK1996,AK1998,STM2001,SH2003,TL2005,SYMT2011,PA2012,DGKK2013,Sotani20b,Sotani21,SK21}. This technique is also useful for the hot neutron stars undergoing thermal evolution \cite{KHA15,SD22,SD24}, as well as to protoneutron stars formed shortly after the core-collapse supernova explosions, e.g., ~\cite{FMP2003,FKAO2015,ST2016,ST2020a,SKTK2017,MRBV2018,SKTK2019,TCPOF19,SS2019,ST2020,STT2021,SMT24}

On the other hand, the presence of dark matter, if any, can also modify the structure of neutron stars as well as their stellar rotation and magnetic field properties. Dark matter admixed neutron stars have been actively studied in the context of both stellar structures and dynamical evolution, assuming various dark matter models. Among several dark matter models, in this study, we focus on the self-interacting fermionic dark matter \cite{Nelson19,Ivanytskyi20,SM24,Rutherford24,KGS25}. Since the dark matter within this model is coupled with the normal matter via only gravitational interaction, the structure and dynamics of the neutron stars should be discussed in the two-fluid formalism, where dark matter and normal matter are modeled as distinct fluids. In this formalism, Tolman-Oppenheimer-Volkof (TOV) equations are modified (see Eqs. (\ref{eq:dm}) -- (\ref{eq:dphi})) \cite{Leung11,Goldman13,Sagun23}, with the total metric potential sourced by both fluids. In addition, the tidal deformability~\cite{Leung22,Das22} and radial oscillations~\cite{Leung12,KS25} for such dark matter admixed neutron stars have been discussed. 

Nevertheless, the analysis of non-radial oscillations in the dark matter admixed neutron star based on the two-fluid formalism remains limited \footnote{Neutron star oscillations within the two-fluid formalism has been discussed in the context of the neutron superfluidity, dealing with the superfluid neutrons as a different kind of fluid from the charged fluid, which leads to an additional mode excitation associated with the superfluidity~\cite{Comer99,Comer02,Andersson02}.}. An exception is the recent study based on the Cowling approximation, which neglects metric perturbations~\cite{SK25}. That work demonstrated the existence of a new oscillation mode associated with dark matter, by solving the perturbation equations derived within the two-fluid framework. However, in that study, the dark matter admixed neutron star models were constructed by fixing the central energy density ratio of dark matter to normal matter, using one specific EOS, i.e., QMC-RMF4, where we could not examine the dependence on the nuclear EOS. In contrast, in this study, we examine the oscillation frequencies of dark matter admixed neutron star models with the fixed dark matter mass fraction, adopting several different EOSs for normal matter component. We then investigate how the resulting frequencies depend on, or are independent of, the nuclear EOS, particularly in the context of universal relations. Specifically, we will focus on the well-established universal relations, which are independent of the EOS uncertainties, discovered in the standard (single-fluid) neutron star models without dark matter, such as
\begin{enumerate}
    \item the relation between compactness, $M/R$, and the dimensionless tidal deformability, $\Lambda_t$;
    \item the relation between the mass-scaled $f$-mode frequency, $f_f M$, and $M/R$; and
    \item the relation between $f_f M$ and $\Lambda_t$.
\end{enumerate}
The relation between $\Lambda_t$ and $M/R$ has already been discussed \cite{Thakur24,Liu24}, but we also confirm it again here.
Our goal is to determine whether these relations persist even in the presence of two-fluid dark matter admixed neutron stars. If the universality of any of these relations breaks down, it may serve as a potential signature to distinguish the dark matter admixed neutron stars from ordinary ones. 
On the other hand, if the universality is still held even with the dark matter, it is also valuable to discuss the properties independently of not only the nuclear EOS but also the dark matter parameters.

This manuscript is organized as follows. In Sec. \ref{sec:NS_multi}, we briefly describe the dark matter model adopted in this study and construct equilibrium configurations of dark matter admixed neutron stars by solving the TOV equations for a two-fluid system. We also present the dimensionless tidal deformability in a two-fluid system. In Sec. \ref{sec:perturbation}, we compute the oscillation frequencies of the stellar models by solving the eigenvalue problem, and examine the corresponding universal relations. Finally, we summarize our findings in this study and present our conclusions in Sec. \ref{sec:Conclusion}. Unless otherwise mentioned, we adopt geometric units with $c=G=1$, where $c$ and $G$ denote the speed of light and the gravitational constant, respectively, and we use the metric signature $(-,+,+,+)$.

\section{Neutron star equilibrium models in two-fluid formalism}
\label{sec:NS_multi}

In this study, we consider static and spherically symmetric neutron star models, which are described with the metric given by
\begin{equation}
  ds^2 = -e^{2\Phi}dt^2 + e^{2\Lambda}dr^2 + r^2 (d\theta^2 + \sin^2\theta\ d\phi^2), \label{eq:metric}
\end{equation}
where $\Phi$ and $\Lambda$ are the metric functions that depend only on the radial coordinate, $r$, and $\Lambda$ is directly related to the mass function, $m$, via $e^{-2\Lambda}=1-2m/r$. We note that $m$ is the gravitational enclosed mass inside $r$. In this study, we consider dark matter admixed neutron stars composed of normal (baryonic) matter and dark matter, which are coupled with each other only via gravitational interaction. The total energy-momentum tensor, $T^{\mu\nu}_{\rm T}$, is expressed as
\begin{equation}
  T^{\mu\nu}_{\rm T} = (\varepsilon_{\rm T} + p_{\rm T})u^\mu u^\nu + p_{\rm T}g^{\mu\nu}, \label{eq:Tmunu0} 
\end{equation}
where $\varepsilon_{\rm T}$, $p_{\rm T}$, and $u^\mu$ are respectively the total energy density, total pressure, and four-velocity of the fluid element. The total energy density and pressure are defined as
\begin{gather}
  \varepsilon_{\rm T} = \sum_x \varepsilon_{x}\ \ {\rm and}\ \   p_{\rm T} = \sum_x p_{x}, \label{eq:epT} 
\end{gather}
using the energy density, $\varepsilon_{x}$, and pressure, $p_{x}$, of the fluid $x$. In this study, we especially consider the two-fluid system, i.e., $x$ denotes either normal matter (NM) or dark matter (DM). The total energy-momentum tensor, $T^{\mu\nu}_{\rm T}$, is decomposed into the individual contributions from each fluid component, $T^{\mu\nu}_{x}$, as
\begin{gather}
  T^{\mu\nu}_{\rm T} = \sum_x T^{\mu\nu}_{x}, \label{eq:Tmunu1} \\
  T^{\mu\nu}_{x} \equiv (\varepsilon_{x} + p_{x})u_{x}^\mu u_{x}^\nu + p_{x}g^{\mu\nu}, \label{eq:Tmunux} 
\end{gather}
where $u^\mu_{x}$ denotes the four-velocity of fluid $x$. From the energy-momentum conservation law ($\nabla_\mu T^{\mu\nu}_{\rm T}=0$) and the additive decomposition in Eq. (\ref{eq:Tmunu1}), it follows that the energy-momentum of each fluid is separately conserved, i.e. $\nabla_\mu T^{\mu\nu}_{x}=0$. 

The TOV equations for a multifluid interacting solely through gravity---governing the stellar structure under static and spherically symmetric conditions---are derived from Einstein equations~\cite{Goldman13,Sagun23}. These equations are explicitly written as
\begin{align}
   m_x' &= 4\pi r^2\varepsilon_{x}, \label{eq:dm} \\
   p_x' &= -\frac{(4\pi r^3p_{\rm T} + m)(\varepsilon_x + p_x)}{r(r-2m)}, \label{eq:dp} \\
   \Phi' &= \frac{4\pi r^3p_{\rm T} + m}{r(r-2m)}, \label{eq:dphi}
\end{align}
where the prime denotes a derivative with respect to $r$, and x $\in$ \{NM, DM\} in present study. To solve the TOV equations, as usual in the case of single fluid neutron stars, one has to prepare the EOS for both normal and dark matter components (see Secs.~\ref{sec:DMmodel} and \ref{sec:DMNSs} for details). 

In addition to mass and radius, the dimensionless tidal deformability, $\Lambda_t$, is another important quantity that characterizes the neutron stars. In a binary system, neutron stars experience tidal deformations due to the gravitational field of their companion. In particular, the tidal deformability plays a crucial role in determining the orbital evolution and the point at which the two stars merge. Larger values of $\Lambda_t$ imply that the stars deform more easily and come into contact at larger separations. The tidal deformability with multipole order $\ell=2$ is constrained through the gravitational wave observations of the binary neutron star merger GW170817. $\Lambda_t$ is related to the dimensionless quadrupole tidal Love number $k_2$ as
\begin{equation}
  \Lambda_{t} = \frac{2}{3}k_2{\cal C}^{-5},
\end{equation}
where ${\cal C}$ denotes the stellar compactness defined by ${\cal C}\equiv M/R$. For gravitationally coupled two-fluid system, the Love number $k_{2}$ is given by
\begin{equation}
    k_{2} = \frac{8{\cal C}^{5}}{5{\cal D}} \left(1-2{\cal C}\right)^{2} 
    \left[2+2{\cal C}\left(y_{R}-1\right)-y_{R}\right],
\end{equation}
with ${\cal D}$ defined as
\begin{align}
   {\cal D}=& 2{\cal C}\left[6-3y_{R}+3{\cal C}\left(5y_{R}-8\right)\right]  \nonumber \\
    &+ 4{\cal C}^{3}\left[13-11y_{R}+{\cal C}\left(3y_{R}-2\right)+2{\cal C}^{2}\left(1+y_{R}\right)\right]  \nonumber \\
    &+3\left(1-2{\cal C}\right)^2 \left[2-y_{R} + 2{\cal C}\left(y_{R}-1\right)\right] \ln\left(1-2{\cal C}\right).
\end{align}
Here, $y_R \equiv y(R)$ is the value of the function $y(r)$ evaluated at the stellar surface, $r=R$. The function $y(r)$ is the solution of the differential equation given by
\begin{equation}
  ry' +y^2 + yF + r^2Q=0
\end{equation}
 with the boundary condition of $y(0)=2$, where $F$ and $Q$ are functions of radial coordinate, $r$, given by 
\begin{align}
   F(r) =& \frac{r-4\pi r^{3}\left(\varepsilon_{\rm T}-p_{\rm T}\right)}{r-2m}, \label{eq:ff} \\
   Q(r) =& \frac{4\pi r^2\left[5\varepsilon_{\rm T} + 9p_{\rm T}  + {\cal G}_{\rm NM} + {\cal G}_{\rm DM}\right] -6}{r(r-2m)} \nonumber \\
    &- 4\left[\frac{m + 4\pi r^{3} p_{\rm T}}{r(r-2m)}\right]^{2}, \label{eq:gg}
\end{align}
using ${\cal G}_x$ is defined for each fluid component, $x = \rm{NM,\ DM}$,~\cite{Leung22,Das22} as 
\begin{equation}
    {\cal G}_x = (\varepsilon_x+p_x)\left(\frac{\partial p_x}{\partial \varepsilon_x}\right)^{-1}.
\end{equation}
These equations are almost the same as those for the single-fluid case derived in Refs. \cite{Hinderer08,Hinderer10,Hippert23}. In particular, by replacing $\varepsilon_{\rm T} \rightarrow \varepsilon_{\rm NM}$ and $p_{\rm T} \rightarrow p_{\rm NM}$, and neglecting ${\cal G}_{\rm DM}$ in Eqs.~(\ref{eq:ff}) and (\ref{eq:gg}), the above equations recover the equations for determining the tidal deformability of a single fluid neutron star composed only of normal matter.

\subsection{Dark Matter Model}
\label{sec:DMmodel}

As in our previous studies~\cite{KGS25,SK25}, we focus here on a self-interacting fermionic dark matter model in which the self-interactions are assumed to be mediated by a hidden vector gauge boson, ${\cal V}^\mu$ (see also Refs.~\cite{Nelson19,Ivanytskyi20,SM24,Rutherford24,KGS25}). In this scenario, dark matter is described by a Dirac fermion, $\chi$, which interacts with normal (baryonic) matter only through gravity and not via any Standard Model interactions. The self-interactions within the dark sector, governed by the vector mediator, influence the EOS of dark matter inside neutron stars. The dynamics of the dark sector and its self-interactions are described by the Lagrangian: 
\begin{align}
    {\cal L}_{\rm DM} =& \,\,\bar\chi\,[ \gamma^\mu (i \partial_\mu - g_\chi {\cal V}_\mu ) - m_\chi] \,\chi \nonumber \\
    & -\frac{1}{2} m_{\rm v}^2 {\cal V}^\mu {\cal V}_\mu - \frac{1}{4} Z^{\mu \nu} Z_{\mu \nu} \ ,
    \label{Eq:LDS}
\end{align}
where $g_\chi$ is gauge coupling constant in the dark sector; $m_\chi$ and  $m_{\rm v}$ denote the masses of the dark matter fermion and the vector mediator, respectively. The field strength tensor of the mediator is defined as $Z^{\mu \nu} = \partial^\mu {\cal V}^\nu-\partial^\nu {\cal V}^\mu$. Observations of the Bullet Cluster (1E 0657-56)~\cite{Randall08} have placed astrophysical constraints on the dark matter self-interaction cross section, yielding $\sigma/m_\chi\lesssim 0.1$ cm$^2$/g, which in turn limits the interaction strength via the coupling-to-mass ratio as $g_\chi/m_v\lesssim 0.1$, as illustrated in Figs. 11 and 12 of Ref.~\cite{KGS25}.

Adopting the mean-field approximation \cite{Rutherford24}, the thermodynamic properties of dark matter in neutron stars can be described through the following expressions for the energy density $\varepsilon_{\rm{DM}}$ and pressure $p_{\rm{DM}}$:
\begin{align}
    \varepsilon_{\rm{DM}} &= \frac{2}{\left(2\pi\right)^{3}} \int^{k_{\chi}^{F}}_{0} \sqrt{k^{2}+m_{\chi}^{2}}\, d^{3}k \,+\,  \frac{1}{2} \left(\frac{g_{\chi}}{m_{\rm{v}}}\right)^{2} n_{\chi}^{2}, \label{eq:rhoDM} \\
    p_{\rm{DM}} &= \frac{2}{3\left(2\pi\right)^{3}} \int^{k_{\chi}^{F}}_{0} \frac{k^{2}}{\sqrt{k^{2}+m_{\chi}^{2}}}\, d^{3}k \,+\,  \frac{1}{2} \left(\frac{g_{\chi}}{m_{\rm{v}}}\right)^{2} n_{\chi}^{2}, \label{eq:pDM}
\end{align}
where $k_{\chi}^{\rm F}$ is the Fermi momentum of dark matter particles and $n_\chi$ represents the number density of dark matter, given by 
\begin{equation}
   n_{\chi} = \frac{2}{\left(2\pi\right)^{3}}\int^{k_{\chi}^{\rm F}}_{0} d^{3}k.
\end{equation} 
In Eq.~(\ref{eq:rhoDM}), the first term corresponds to the kinetic and rest-mass energy of dark matter fermions, derived from the standard relativistic energy-momentum relation. The second term accounts for the self-interaction energy mediated by the vector field ${\cal V}^\mu$ and introduces an effective repulsive contribution to the dark matter EOS. Similarly, in Eq.~(\ref{eq:pDM}), the first term represents the Fermi pressure, while the second term arises from the pressure contribution due to self-interactions.

\subsection{Dark Matter admixed Neutron Star Models}
\label{sec:DMNSs}

Before discussing the oscillation frequencies excited in dark matter admixed neutron stars, we first construct the corresponding equilibrium background models. This requires adopting both a normal matter EOS and the dark matter EOS described in the previous subsection. In this study, we focus on four representative nuclear matter EOSs listed in Table~\ref{tab:EOS}. Among them, DD2~\cite{DD2} and QMC-RMF4~\cite{QMC4} are based on the relativistic mean field approximation; SKa~\cite{SKa} is constructed using a Skryme-type effective interaction; and Togashi~\cite{Togashi17} is built via the variational method. In addition, DD2, SKa, and Togashi are unified EOS, i.e., the crustal EOS is expressed with the same frame work as in the core, while QMC-RFM4 is connected to SLy4~\cite{SLy4} in the crustal region.

To model the dark matter sector, one must specify two key parameters, i.e., the interaction strength normalized by the mass of vector mediator, $g_\chi/m_v$, and the dark matter particle mass, $m_\chi$. Following~Ref.~\cite{KGS25}, we adopt the dark matter particle mass to $m_\chi=5$ GeV as a representative value, which is consistent with the cosmological constraints from the Planck observation~\cite{Zurek14}. In particular, the observed ratio of dark matter to baryonic matter densities in the Universe is approximately 5, which suggests that particle masses near the GeV scale are consistent with thermal relic production~\cite{Zurek14}. The interaction strength is varied in the range $0\le g_\chi/m_v\le 0.1$, in accordance with the Bullet Cluster constraints on self-interaction cross sections~\cite{KGS25}. 


The dark matter admixed neutron star models are constructed by specifying the central energy densities of normal matter and dark matter, $\varepsilon_c^{\rm NM}$ and $\varepsilon_c^{\rm DM}$, together with their respective EOS. The resulting stellar configurations can be parametrized in two equivalent ways. First, one can vary $\varepsilon_c^{\rm NM}$ while fixing the ratio $\varepsilon_c^{\rm DM}/\varepsilon_c^{\rm NM}$, as done in Refs.~\cite{KGS25,SK25}. Alternatively, one can characterize the models by the dark matter mass fraction, $M_{\rm DM}/M$, where $M_{\rm{DM}}$ is the gravitational mass contributed by dark matter and $M$ is the total mass of the star. In this study, we adopt the latter approach and construct stellar sequences with fixed dark matter mass fractions.

\begin{table}
\caption{EOS parameters, $K_0$, $L$, and $\eta\equiv(K_0L^2)^{1/3}$ \cite{SIOO14}, for normal matter EOS adopted in this study. The maximum mass of spherically symmetric neutron stars without dark matter are also listed.} 
\label{tab:EOS}
\begin {center}
\renewcommand{\arraystretch}{1.5} 
\setlength{\tabcolsep}{3.5pt}
\begin{tabular}{ccccc}
\hline\hline
EOS & $K_0$ (MeV) & $L$ (MeV) & $\eta$ (MeV) & $M_{\rm max}/M_\odot$    \\ 
\hline
DD2 & 243 & 55.0  & 90.2  & 2.41   \\
QMC-RMF4  &  279  & 31.3  & 64.9  &  2.21  \\
SKa  &  263  & 74.6  & 114  & 2.22   \\
Togashi & 245 & 38.7 & 71.6 & 2.21   \\
\hline \hline
\end{tabular}
\end {center}
\end{table}

\begin{figure*}[tbp]
\begin{center}
\includegraphics[width=0.9\textwidth,height=0.5\textheight,keepaspectratio]{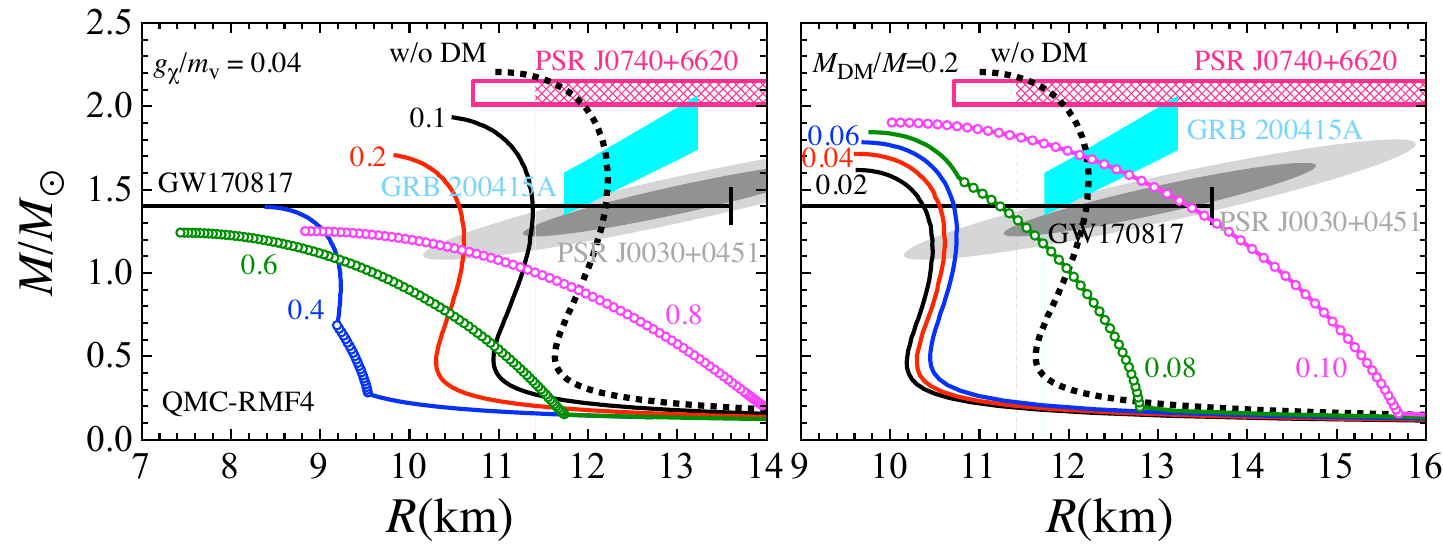}
\end{center}
\caption{
Mass-radius relations for dark matter admixed neutron star models constructed using the QMC-RMF4 EOS for baryonic matter. \textbf{Left panel:} Results for fixed interaction strength $g_\chi/m_v=0.04$ MeV$^{-1}$, with varying dark matter mass fractions $M_{\rm DM}/M=0.1$, 0.2, 0.4, 0.6, and 0.8. \textbf{Right panel:} Results for fixed dark matter mass fraction $M_{\rm DM}/M=0.2$, with varying interaction strengths $g_\chi/m_v=0.02$, 0.04, 0.06, 0.08, and 0.10 MeV$^{-1}$. In both panels, the stellar radius, $R$, is defined as the outermost surface, i.e., $R={\rm max}(R_{\rm NM}, R_{\rm DM})$. Solid lines correspond to dark-core configurations where $R_{\rm DM}<R_{\rm NM}$, while open circles represent dark-halo configurations with$R_{\rm DM}>R_{\rm NM}$. For reference, the mass-radius relation for neutron star models without dark matter is also shown with the dotted line, and several astronomical constraints from PSR J0740+6620, PSR J0030+0451, GW170817, and GRB 200415A are also included (see text for details).
}
\label{fig:MR1}
\end{figure*}

In Fig.~\ref{fig:MR1}, we present the mass-radius relations for dark matter admixed neutron star models constructed with QMC-RMF4 as the baryonic EOS. The left panel shows sequences with fixed interaction strength $g_\chi/m_v=0.04$ MeV$^{-1}$ for varying dark matter mass fractions, i.e., $M_{\rm DM}/M=0.1$, 0.2, 0.4, 0.6, and 0.8. In contrast, the right panel depicts sequences with fixed dark matter mass fraction $M_{\rm DM}/M=0.2$ for different interaction strengths, i.e., $g_\chi/m_v=0.02$, 0.04, 0.06, 0.08, and 0.10 MeV$^{-1}$. In both panels, solid lines represent dark-core configurations, where the dark matter radius is smaller than that of normal matter ($R_{\rm DM}<R_{\rm NM}$), while the open circles denote dark-halo configurations with $R_{\rm DM}>R_{\rm NM}$. From this figure, it is evident that for a fixed interaction strength, the maximum mass of the stellar models decreases as the dark matter mass fraction increases. Conversely, for a fixed dark matter mass fraction, the maximum mass increases with strong interaction strength. While the maximum masses of some of these dark matter admixed models fall below the observed $2M_{\odot}$ mass constraint, we intentionally explore a broad parameter space in both dark matter mass fraction and interaction strength to investigate potential universal behavior, as discussed in the subsequent sections.

On the other hand, as discussed in Refs.~\cite{Yagi14,GGRB21} for neutron stars without dark matter, the dimensionless tidal deformability is known to correlate strongly with the stellar compactness, exhibiting only weak dependence on the underlying EOS. To verify this behavior across different EOS, we derive an empirical relation between the dimensionless tidal deformability and stellar compactness for neutron star models without dark matter, given by
\begin{align}
  \log_{10}(\Lambda_t) =& 0.0018274/\tilde{\cal C} + 9.7040 - 9.9870\tilde{\cal C}^{1/2} \nonumber \\ 
  & + 3.4177\tilde{\cal C} -0.5743\tilde{\cal C}^2,
  \label{eq:C_Lam}
\end{align}
where $\tilde{\cal C}={\cal C}/0.172$~\cite{SK25} is the normalized compactness (see Appendix for details). Here, 0.172 corresponds to the compactness of a canonical stellar model with mass $1.4M_\odot$ and radius 12 km. This relation reproduces the dimensionless tidal deformability to within $\sim 20\%$ accuracy for canonical and massive neutron stars without dark matter (see Fig.~\ref{fig:C-L}).

\begin{figure*}[tbp]
\begin{center}
\includegraphics[width=\textwidth,height=0.5\textheight,keepaspectratio]{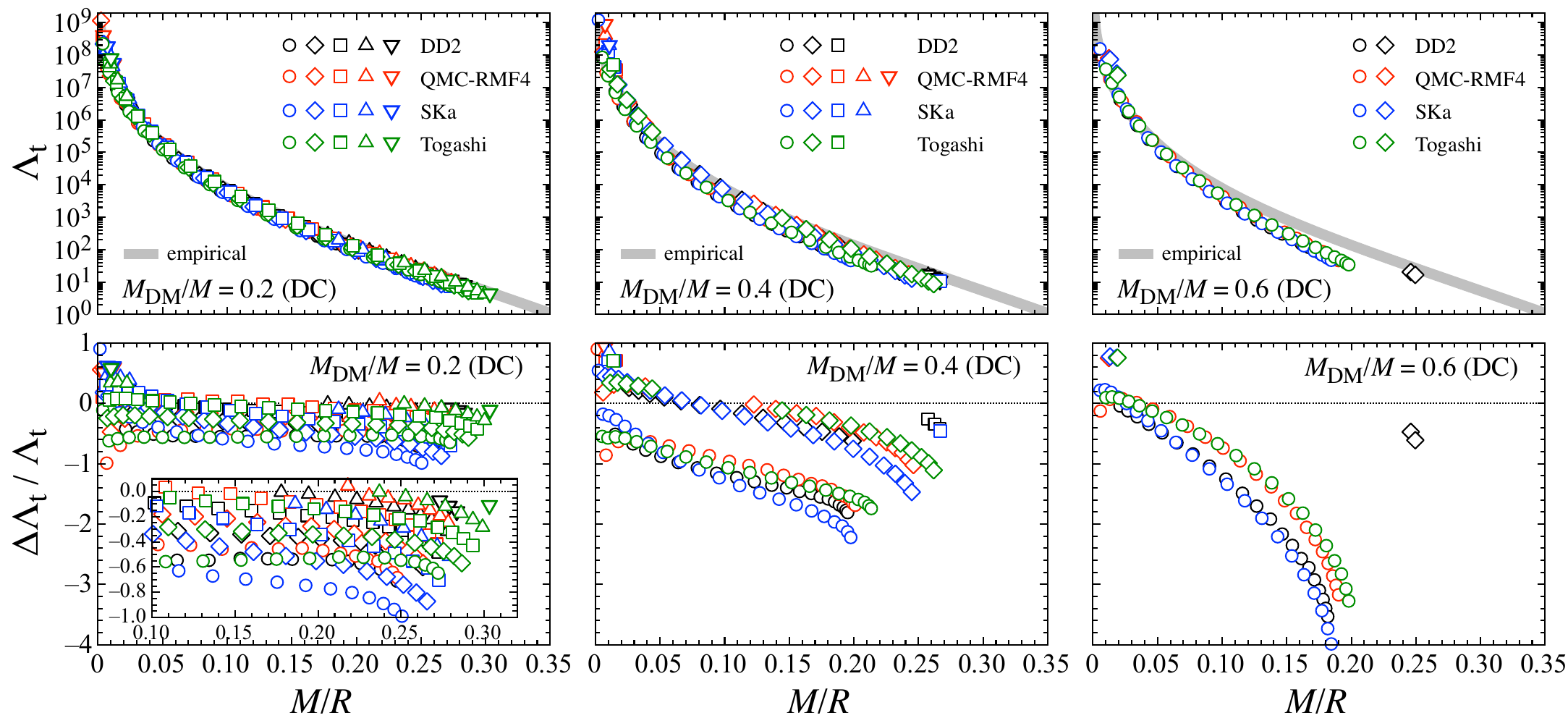}
\end{center}
\caption{
Top panels: Relation between the dimensionless tidal deformability ($\Lambda_t$) and stellar compactness ($M/R$) for various dark matter admixed neutron star models with dark core configurations. The left, middle, and right panels correspond to stellar models with fixed dark matter mass fractions of $M_{\rm DM}/M=0.2$, 0.4, and 0.6, respectively. Different interaction strengths are indicated by symbols: circles $(g_\chi/m_v=0.02\ \rm{MeV}^{-1})$, diamonds $(g_\chi/m_v=0.04\ \rm{MeV}^{-1})$, squares $(0.06\ \rm{MeV}^{-1})$, triangles $(0.08\ \rm{MeV}^{-1})$, and inverted triangles $(0.10\ \rm{MeV}^{-1})$. The thick solid line in each panel represents the empirical relation between $\Lambda_t$ and $M/R$ for standard neutron star models without dark matter, as given by Eq.~(\ref{eq:C_Lam}). Bottom panels: Relative deviation of the calculated $\Lambda_t$ from the empirical estimate $\Lambda_t^{\rm em}$, defined as $\Delta \Lambda_t/\Lambda_t=(\Lambda_t^{\rm em}-\Lambda_t)/\Lambda_t$. An enlarged view of this deviation for $M_{\rm DM}/M=0.2$ is shown in the inset of the bottom-left panel.
}
\label{fig:DC-C-Lam}
\end{figure*}

\begin{figure*}[tbp]
\begin{center}
\includegraphics[width=\textwidth,height=0.5\textheight,keepaspectratio]{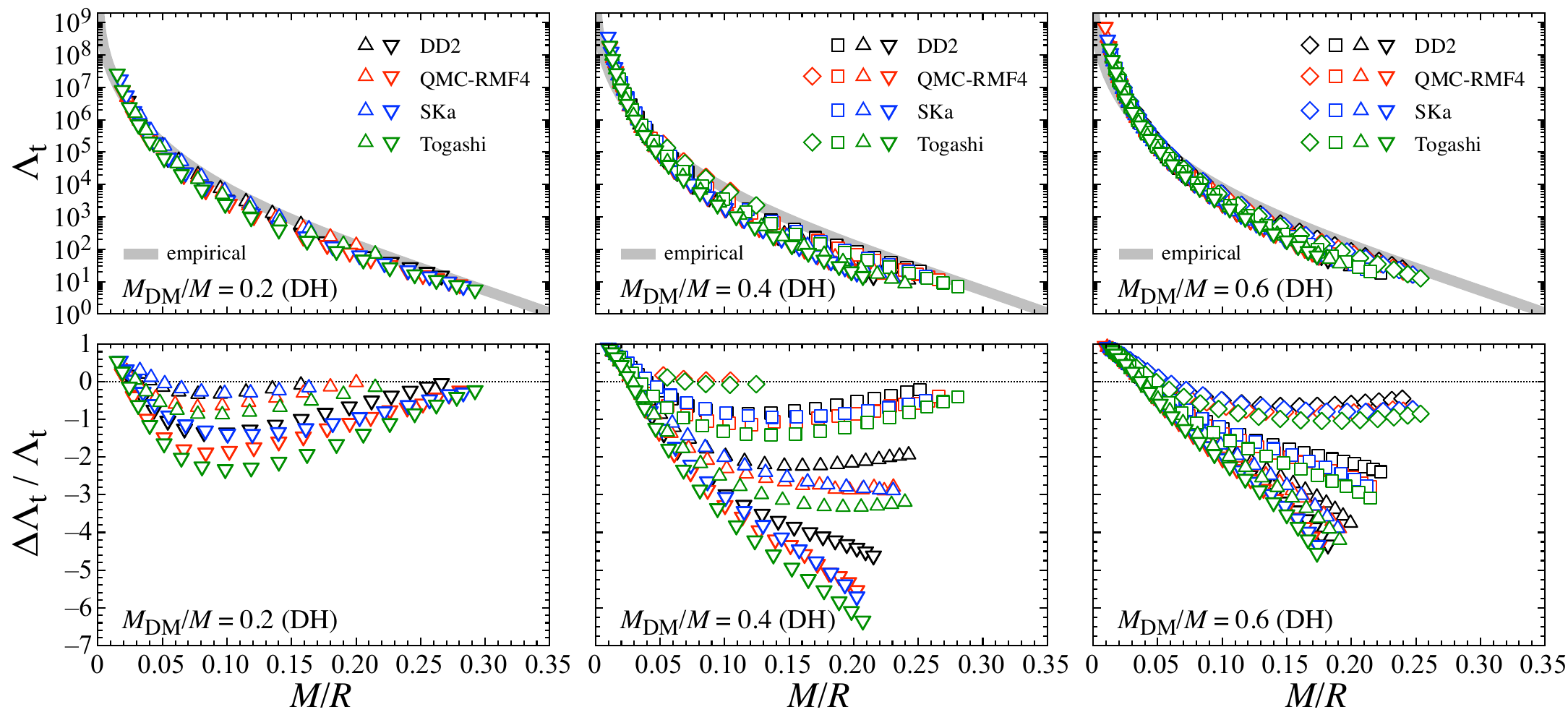} 
\end{center}
\caption{
Same as Fig.~\ref{fig:DC-C-Lam}, but for dark halo configurations.
}
\label{fig:DH-C-Lam}
\end{figure*}

For dark matter admixed neutron star models, we present the dimensionless tidal deformability as a function of the stellar compactness, $M/R$, with the total gravitational mass, $M$, and $R$ denotes the outermost surface radius of the two-fluid configuration. Figure~\ref{fig:DC-C-Lam} shows the results for dark-core configurations, while Fig.\ref{fig:DH-C-Lam} displays those for dark-halo configurations, using the four baryonic EOS listed in Table~\ref{tab:EOS}. In both figures, the circles, diamonds, squares, triangles, and inverted triangles correspond to $g_\chi/m_v = 0.02$, 0.04, 0.06, 0.08, and 0.10 MeV$^{-1}$, respectively, and the thick solid line represents the empirical $\Lambda_t-{\cal C}$ relation for standard neutron stars without dark matter, given by Eq.~(\ref{eq:C_Lam}). The bottom panels additionally display the relative deviation, defined as
\begin{align}
    \Delta \Lambda_t \equiv \frac{(\Lambda_t^{\rm em} - \Lambda_t)} {\Lambda_t},
\end{align}
where $\Lambda_t^{\rm em}$ is the tidal deformability estimated from Eq.~(\ref{eq:C_Lam}) and $\Lambda_t$ is the actual value computed for the dark matter admixed neutron star. From these figures, it is evident that the dimensionless tidal deformability for dark matter admixed neutron star models significantly deviates from the empirical relation for standard neutron stars without dark matter given by Eq.~(\ref{eq:C_Lam}). Moreover, in Fig.\ref{fig:DC-C-Lam}, the deviation from the empirical relation becomes more pronounced as the interaction strength decreases for fixed dark matter fraction in dark-core configurations. Conversely, in Fig.~\ref{fig:DH-C-Lam}, the deviation increases with interaction strength for dark-halo configurations at fixed dark matter fraction. Therefore, simultaneous observation of stellar compactness and dimensionless tidal deformability could help distinguish dark matter admixed neutron stars from ordinary neutron stars without dark matter.

\section{Oscillation frequencies}
\label{sec:perturbation}

As in our previous study~\cite{SK25}, we investigate the oscillation frequencies, assuming the Cowling approximation\footnote{It is known that the frequencies calculated with the Cowling approximation systematically deviate at most $\sim 30\%$ from those determined with the metric perturbations, at least, for ordinary neutron stars without dark matter. In this study, we simply adopt the Cowling approximation, while it should be checked whether our results work for the case with metric perturbations. Anyway, parameter studies remain troublesome, even if the calculations with metric perturbations become possible. Discussing universality with the Cowling approximation first serves as crucial preparation for performing the calculations with meric perturbations.}. Since the dark matter is assumed to couple with normal matter only through gravity, the linearized perturbation equations for each fluid---derived from the energy-momentum conservation law---become decoupled. In practice, the perturbation equations are expressed in terms of the Lagrangian displacement in the radial and angular directions for fluid $x$, denoted by $W_x$ and $V_x$, respectively:
\begin{gather}
    W'_x  = \frac{1}{c_{s,x}^2}\left[\Phi' W_x +\omega_x^2 r^2 e^{-2\Phi+\Lambda}V_x\right] - \ell(\ell+1)e^{\Lambda}V_x,  \label{eq:dW} \\
   V'_x = -e^{\Lambda}\frac{W_x}{r^2} + 2\Phi' V_x  
       - {\cal A}_x\left[\frac{\Phi'}{\omega_x^2r^2} e^{2\Phi-\Lambda}W_x + V_x\right], \label{eq:dV}
\end{gather}
where $\ell$ is the azimuthal quantum number, $\omega_x$ is the oscillation (real) eigenvalue for the fluid $x$, $c_{s,x}$ is the sound speed of the fluid $x$, and ${\cal A}_x$ is the relativistic Schwarzschild discriminant. These quantities are defined as
\begin{align}
  c_{s,x}^2 &\equiv \frac{\partial p_x}{\partial \varepsilon_x}\bigg|_s = \frac{\gamma_x p_x}{\varepsilon_x + p_x}, \\
  {\cal A}_x(r) &= \frac{1}{\varepsilon_x + p_x}  \left(\varepsilon_x' - \frac{ p_x'}{c_{s,x}^2}\right),
\end{align}
where $\gamma_x$ is the adiabatic index of fluid $x$. Once $\omega_x$ is determined, the corresponding eigenfrequency is given by $f_x=\omega_x/(2\pi)$. The appropriate boundary conditions to integrate these perturbation equations consist of regularity conditions at the stellar center and the requirement that the Lagrangian pressure perturbation should be zero at the surface of each fluid component. In this study, we focus on the $\ell=2$ modes, as they are expected to dominate the gravitational-wave signal. In addition, we simply assume that $c_{s,x}^2=dp_x/d\varepsilon_x$ as in our previous study \cite{SK25}, where only the fundamental ($f$-) and pressure ($p$-) modes can be excited.

\begin{figure}[tbp]
\begin{center}
\includegraphics[width=0.45\textwidth,height=0.5\textheight,keepaspectratio]{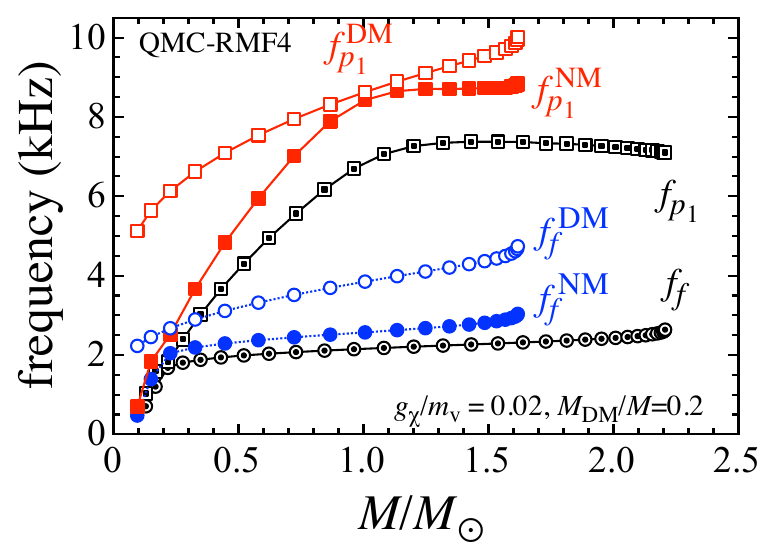} 
\end{center}
\caption{
The $f$- and $p_1$-mode frequencies associated with normal (dark) matter, $f_f^{\rm NM}$ ($f_f^{\rm DM}$) and $f_{p_1}^{\rm NM}$ ($f_{p_1}^{\rm DM}$) are shown as a function of the total mass for the dark matter admixed neutron stars with a fixed DM mass fraction of $M_{\rm DM}/M=0.2$ and interaction strength $g_\chi/m_v=0.02$, adopting QMC-RMF4 as normal matter EOS. For reference, the $f$- and $p_1$-mode frequencies of standard neutron stars without dark matter are also shown, denoted by double-circles and double-squares markers. 
}
\label{fig:Mf-g002-DMF020-QMC}
\end{figure}

In Fig.~\ref{fig:Mf-g002-DMF020-QMC}, we present the $f$-mode and first pressure ($p_1$-mode) oscillation frequencies associated with the normal matter component---denoted $f_f^{\rm NM}$ and $f_{p_1}^{\rm NM}$---as well as those associated with the dark matter component--- denoted $f_f^{\rm DM}$ and $f_{p_1}^{\rm DM}$. These are shown as a function of the total stellar mass for dark matter admixed neutron star models constructed with $g_\chi/m_v=0.02\ \rm{MeV}^{-1}$ and a fixed dark matter mass fraction $M_{\rm DM}/M=0.2$, adopting QMC-RMF4 as normal matter EOS. Filled markers correspond to the frequencies associated with normal matter, while open markers denote those of the dark matter component. For comparison, the oscillation frequencies of neutron star models without dark matter—constructed with the same baryonic EOS—are also shown, using double-circle and double-square markers for the $f$- and $f_{p_{1}}$-modes, respectively. From the right panel of Fig.~\ref{fig:MR1}, one can see that the corresponding dark matter admixed neutron star configurations (shown with a black solid line) exhibit a dark-core structure. These stars are systematically more compact than their dark matter-free counterparts, as seen by comparing the solid and dotted lines in Fig.~\ref{fig:MR1}. As a result of this increased compactness, the oscillation frequencies of the dark matter admixed neutron stars tend to be higher than those of the standard neutron stars without dark matter. 

\begin{figure*}[tbp]
\begin{center}
\includegraphics[width=\textwidth,height=0.5\textheight,keepaspectratio]{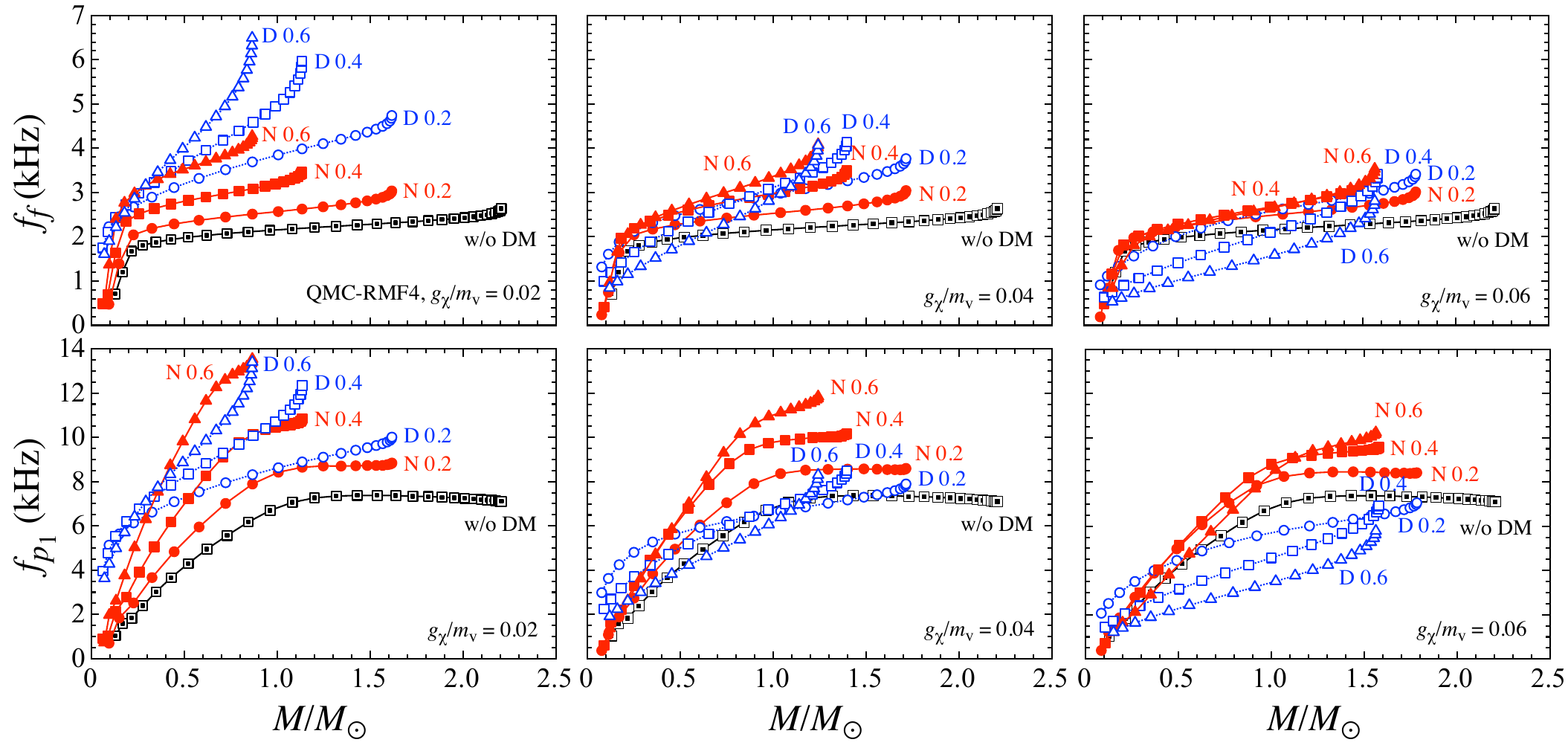} 
\end{center}
\caption{
The $f$- and $p_1$-mode frequencies are shown in the top and bottom panels, respectively, as a function of the total stellar mass for dark matter admixed neutron stars with three different dark matter mass fractions: $M_{\rm DM}/M = 0.2$, 0.4, and 0.6. The results are computed using QMC-RMF4 as the EOS for normal matter. The left, middle, and right panels correspond to interaction strengths of $g_\chi/m_v = 0.02$, 0.04, and 0.06 MeV$^{-1}$, respectively. In each panel, mode frequencies associated with normal (dark) matter are labeled as N~0.2, N~0.4, N~0.6 (D~0.2, D~0.4, D~0.6), where 0.2, 0.4, and 0.6 denote adopted dark matter mass fractions. For reference, the $f$- and $p_1$-mode frequencies excited in the neutron star models without dark matter are also shown with double-square markers. 
}
\label{fig:Mf-QMC}
\end{figure*}

In Fig.~\ref{fig:Mf-QMC}, the $f$- and $p_1$-mode oscillation frequencies for dark matter admixed neutron stars constructed using QMC-RMF4 as normal matter EOS, adopting various combinations of dark matter mass fractions and interaction strengths, are shown in the top and bottom panels, respectively. The left, middle, and right panels correspond to interaction strengths of $g_\chi/m_v=0.02$, 0.04, and 0.06 MeV$^{-1}$, respectively. In each panel, filled markers (circles, squares, and diamonds) denote the oscillation frequencies associated with the normal matter component, while open markers represent those associated with the dark matter component. In both cases, different marker shapes---circles, squares, and diamonds---indicate different dark matter mass fractions: $M_{\rm{DM}}/M$=0.2, 0.4, and 0.6. We note that we systematically did the parameter study in a relatively wider parameter space, although the adopted parameters are not so dense. So, we believe that our results shown in this study is reliable, at least within the parameter space we adopted. For reference, the corresponding mode frequencies of standard neutron stars without dark matter, constructed using the same baryonic EOS, i.e, QMC-RMF4, are also shown in both panels. From this figure, one can observe a general trend: for configurations with a dark core structure, the oscillation frequencies associated with the normal matter are lower than those associated with dark matter. Conversely, in configurations exhibiting a dark halo structure, the frequencies of the normal matter tend to be higher than those of the dark matter component. This behavior reflects the underlying differences in stellar average density and fluid stratification between the two configurations. 

\begin{figure*}[tbp]
\begin{center}
\includegraphics[width=0.9\textwidth,height=0.4\textheight,keepaspectratio]{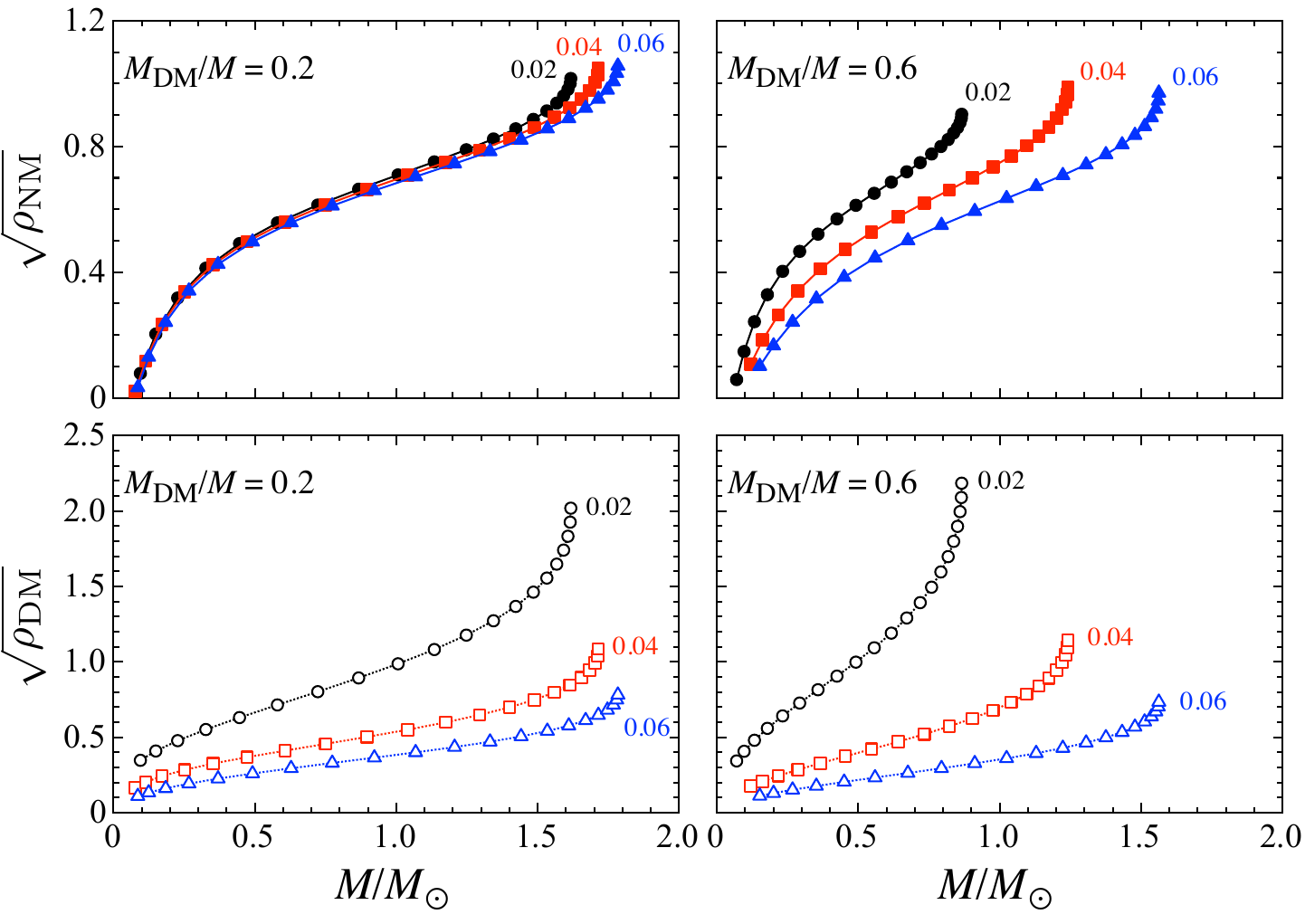} 
\end{center}
\caption{
The square root of the average density for normal matter, $\sqrt{\rho_{\rm NM}}$, and dark matter, $\sqrt{\rho_{\rm DM}}$, is shown in the top and bottom panels, respectively, as a function of total stellar mass for $M_{\rm DM}/M=0.2$. The average densities are defined as defined as $\rho_{\rm NM}\equiv (M_{\rm NM}/1.4M_\odot) (R_{\rm NM}/10\ {\rm km})^{-3}$ and $\rho_{\rm DM}\equiv (M_{\rm DM}/1.4M_\odot)(R_{\rm DM}/10\ {\rm km})^{-3}$. Results are shown for dark matter mass fractions of 0.2 (left panel) and 0.6 (right panel), using QMC-RMF4 as the equation of state for normal matter. In each panel, circles, squares, and triangles correspond to $g_\chi/m_v = 0.02$, 0.04, and 0.06, respectively.
}
\label{fig:M_ave}
\end{figure*}

The behavior of the $f$- and $p_1$-mode frequencies can be qualitatively understood in terms of the stellar average densities of the normal and dark matter components. We define the normalized average densities as
\begin{gather}
  \rho_{\rm NM} = \frac{M_{\rm NM}}{1.4M_\odot}\left(\frac{R_{\rm NM}}{10\ {\rm km}}\right)^{-3}, \\
  \rho_{\rm DM} = \frac{M_{\rm DM}}{1.4M_\odot}\left(\frac{R_{\rm DM}}{10\ {\rm km}}\right)^{-3}, 
\end{gather}
where $M_{\rm{NM}}$ and $M_{\rm{DM}}$ denote the total mass of the normal and dark matter fluids, and $R_{\rm{NM}}$ and $R_{\rm{DM}}$ correspond to their respective radii. Both densities are normalized by the canonical mass ($1.4M_\odot$) and radius of $10$ km. This formulation is motivated by the fact that $f$- and $p_1$-modes are predominantly acoustic in nature, and their frequencies are generally characterized by the sound speed, which scales as the square root of the average density, i.e., $\sqrt{\rho_{\rm NM}}$ and $\sqrt{\rho_{\rm DM}}$, respectively. In Fig.~\ref{fig:M_ave}, we show $\sqrt{\rho_{\rm NM}}$ (top panels) and $\sqrt{\rho_{\rm DM}}$ (bottom panels) as functions of the total gravitational mass. The left and right panels correspond to mass fractions of $M_{\rm DM}/M=0.2$ and 0.6, respectively. In each panel, different marker shapes represent different interaction strengths: circles, squares, and triangles correspond to $g_\chi/m_v=0.02$, 0.04, and 0.06 MeV$^{-1}$, respectively. From the left-top panel, it is evident that for $M_{\rm DM}/M=0.2$, the average density of normal matter component, $\rho_{\rm{NM}}$, remains nearly independent of the interaction strength. This trend aligns with the nearly constant behavior of $f$- and $p_1$-mode frequencies associated with normal matter (labeled as N0.2) in Fig.~\ref{fig:Mf-QMC}, where the dependence on $g_{\chi}/m_{v}$ is minimal. In contrast, for $M_{\rm{DM}}/M = 0.6$, we observe that both $\sqrt{\rho_{\rm{NM}}}$ and $\sqrt{\rho_{\rm{DM}}}$ decrease with increasing interaction strength (Fig.~\ref{fig:M_ave}). A similar trend is also seen for $\sqrt{\rho_{\rm{DM}}}$ at $M_{\rm{DM}}/M = 0.2$. These systematic variations are reflected in Fig.~\ref{fig:Mf-QMC}, where $f_{f}^{\rm{NM}}$ and $f_{p_{1}}^{\rm{NM}}$ (N0.6), as well as $f_{f}^{\rm{DM}}$ and $f_{p_{1}}^{\rm{DM}}$ (D0.2 and D0.6), decrease from the left to the right panels as $g_{\chi}/m_v$ increases. Nevertheless, despite these trends, we find that the oscillation frequencies $f_{f}^{\rm{NM}}$ and $f_{f}^{\rm{DM}}$ cannot be universally expressed as functions of $\sqrt{\rho_{\rm{NM}}}$ or $\sqrt{\rho_{\rm{DM}}}$. This suggests that while the average density captures part of the frequency behavior, additional structural details and fluid interactions play a role in determining the precise oscillation spectrum.

We now turn to the examination of universal relations involving the mass-scaled $f$-mode frequencies in terms of two key global quantities: the stellar compactness and the dimensionless tidal deformability. Such relations are particularly valuable for inferring neutron star properties from gravitational wave and astroseismic observations, as they are expected to be largely independent of the underlying EOS. For the standard neutron star stars without dark matter, the mass-scaled $f$-mode frequency exhibits a robust empirical relation with the normalized stellar compactness, $\tilde{\cal C}$. This relation takes the form:
\begin{align}
   f_f M\ ({\rm kHz}/M_\odot) =&  -0.01932 + 2.115\tilde{\cal C} + 1.731\tilde{\cal C}^2 \nonumber \\
   & -0.5720\tilde{\cal C}^3, \label{eq:ffM_C}
\end{align}
as established in Ref.~\cite{SK25}. This relation enables estimation of the $f$-mode frequency with an accuracy of approximately 1\% for canonical and massive neutron star models (i.e., $\tilde{\cal C}\gtrsim 1$, or equivalently $M/R\gtrsim 0.17$) in the absence of dark matter. In parallel, a complementary universal relation between the mass-scaled $f$-mode frequency (calculated with the Cowling approximation) and the dimensionless tidal deformability, $\Lambda_t$, is derived in Appendix~\ref{sec:appendix_1} as the following fitting formula
\begin{align}
  f_f M& \ ({\rm kHz}/M_\odot ) = 6.4012-0.2049\chi-0.8612\chi^2 \nonumber \\ 
  &+0.2376\chi^3-0.025063\chi^4+0.00095864\chi^5,
  \label{eq:ffM_Lam}
\end{align}
where $\chi=\log_{10}\Lambda_t$. Using this relation, one can estimate the $f$-mode frequencies for canonical and massive standard DM-free neutron stars ($\Lambda_t\lesssim 1000$) with a typical accuracy of a few percent.

\begin{figure}[tbp]
\begin{center}
\includegraphics[scale=0.58]{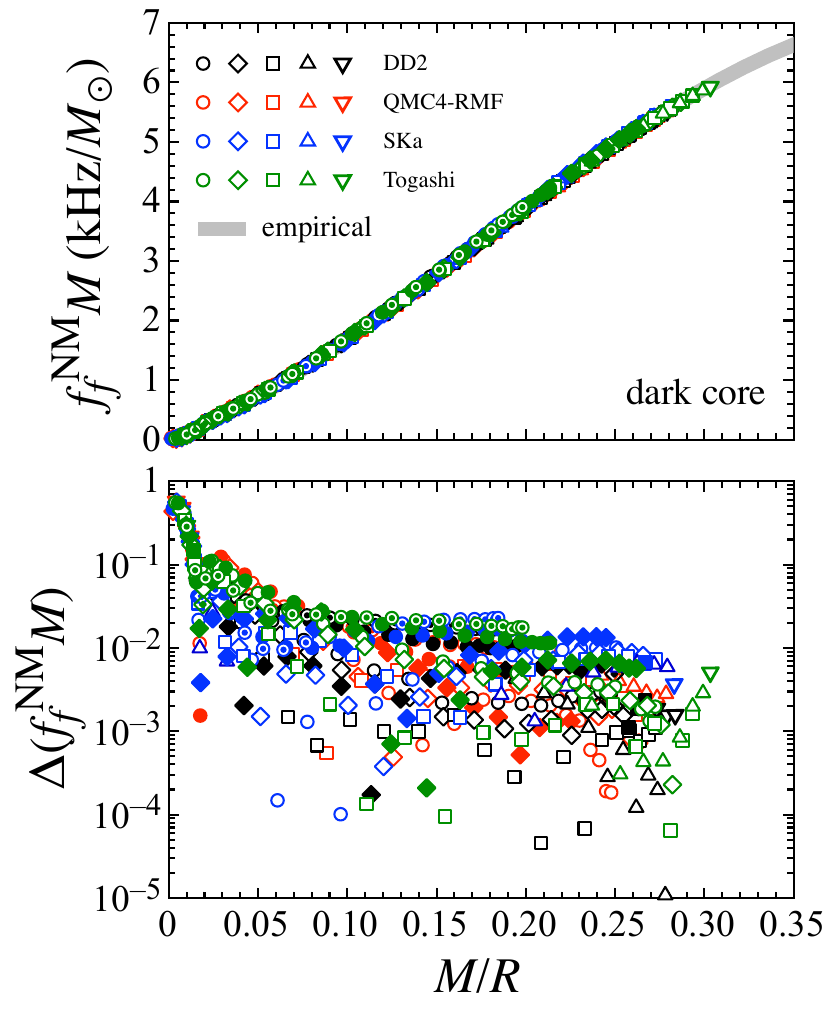} 
\end{center}
\caption{
Mass-scaled $f$-mode frequencies associated with normal matter ($f_f M$) are shown as a function of stellar compactness for dark core configurations, constructed using various nuclear EOSs, interaction strengths, and dark matter mass fractions. Circles, diamonds, squares, triangles, and inverted triangles correspond to $g_\chi/m_v = 0.02$, 0.04, 0.06, 0.08, and 0.10 MeV$^{-1}$, respectively, while open, filled, and double symbols represent $M_{\rm DM}/M = 0.2$, 0.4, and 0.6. The thick solid line shows the empirical relation for neutron stars without dark matter, given by Eq.~(\ref{eq:ffM_C}). The bottom panel displays the relative deviation of $f_f M$ from the empirical prediction, computed using Eq.~(\ref{eq:relative}). 
}
\label{fig:ffM_D-MR-DC}
\end{figure}

\begin{figure}[tbp]
\begin{center}
\includegraphics[scale=0.58]{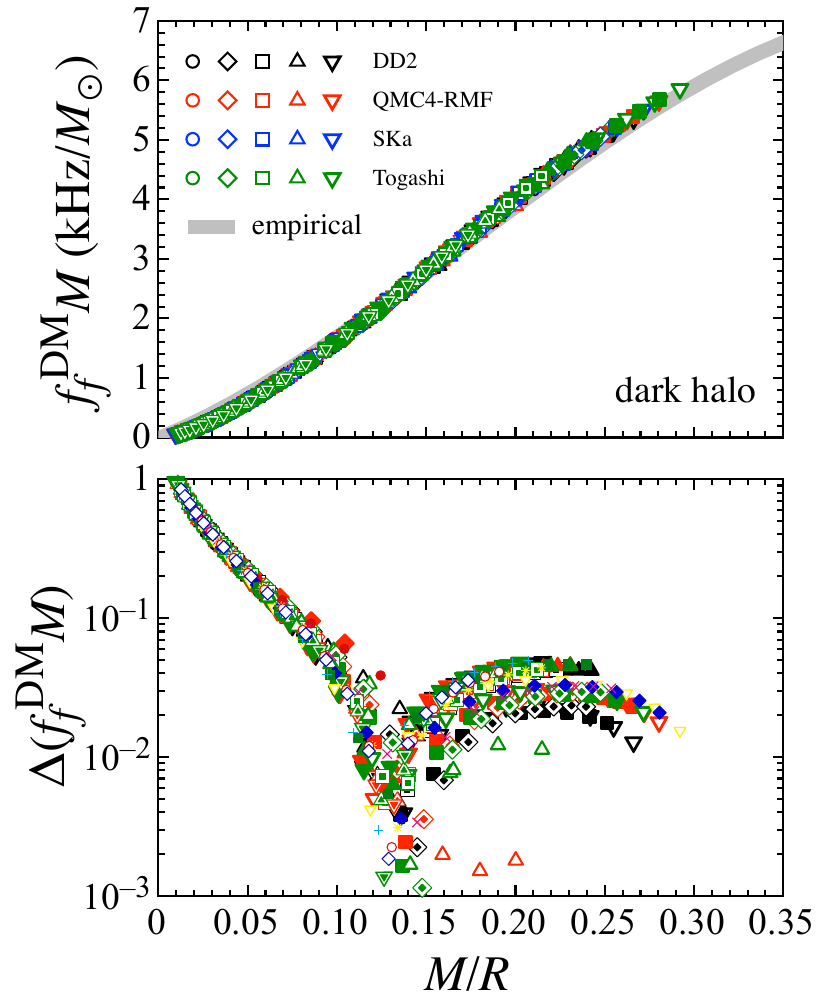} 
\end{center}
\caption{
Same as Fig.~\ref{fig:ffM_D-MR-DC}, but focusing on the dark halo configurations.
}
\label{fig:ffM_D-MR-DH}
\end{figure}

In Fig.~\ref{fig:ffM_D-MR-DC}, we present the mass-scaled $f$-mode frequencies associated with normal matter from the dark core configurations constructed using various nuclear matter EOSs, dark matter mass fraction, and interaction strengths. These results are compared against the empirical relation given by Eq.(\ref{eq:ffM_C}). Similarly, Fig.~\ref{fig:ffM_D-MR-DH} shows the corresponding mass-scaled $f$-mode frequencies associated with dark matter for dark halo configurations, using the same set of model parameters. In both figures, the lower panels display the relative deviation of the numerical result from the empirical prediction, defined as
\begin{equation}
  \Delta(A) \equiv \frac{|A^{\rm N}-A^{\rm E}|}{A^{\rm N}} \label{eq:relative}
\end{equation}
where $A^{\rm N}$ denotes the value obtained from numerical simulations, and $A^{\rm E}$ represents the value estimated using the empirical fit. From these figures, focusing on the canonical and massive neutron star models ($M/R\gtrsim 0.17$), we find that the mass-scaled $f$-mode frequencies associated with normal matter in dark core configuration and those associated with dark matter in dark halo configurations, i.e., the mass-scaled $f$-mode frequencies associated with the fluid whose radius exceeds that of the other component, can be well described as a function of (total) stellar compactness, $M/R$. This universality holds independently of the nuclear matter EOS and the parameters of the dark matter model adopted in this study, and is consistent with the behavior found for neutron star models without dark matter\footnote{Although this feature was previously demonstrated in Ref.~\cite{SK25} for models constructed with a fixed central density ratio of dark matter to normal matter, and adopting only the QMC-RMF4 as the nuclear EOS, the present work is the first to establish the universality across multiple EOSs and a wide range of dark matter parameters, for stellar sequences constructed at fixed dark matter mass fraction.}\footnote{We confirmed that the universality shown in Figs.~\ref{fig:ffM_D-MR-DC} and \ref{fig:ffM_D-MR-DH} are held even with two additional EOS, i.e., SkI3 and SLy4. But, since Figs.~\ref{fig:ffM_D-MR-DC} and \ref{fig:ffM_D-MR-DH} are already messy, we did not show them here.}. By contrast, we also explored whether a universal relation exists for the $f$-mode frequencies associated with dark matter in dark core configurations and those with normal matter in dark halo configurations, but no such universal relation was found (or could be identified) in our results.

Meanwhile, in Fig.~\ref{fig:ffMD-Lam-DC}, we present the mass-scaled $f$-mode frequencies associated with normal matter in dark-core configurations, constructed using various EOSs for normal matter and different dark matter parameters, plotted as a function of the dimensionless tidal deformability. {In these plots, different marker shapes—circles, diamonds, squares, triangles, and inverted triangles—represent interaction strengths of $g_\chi/m_v=0.02$, 0.04, 0.06, 0.08, 0.10 MeV$^{-1}$, respectively. Additionally, open, filled, and double-marked symbols correspond to dark matter mass fractions of $M_{\rm DM}/M=0.2$, 0.4, and 0.6, respectively. For comparison, the empirical relation for neutron star models without dark matter is shown as a thick solid line. The bottom panel displays the relative deviation of the numerical results ($f_f^{\rm NM}M$) from this empirical fit, highlighting the extent to which the dark-core configurations deviate from the DM-free universal trend. 

Similarly, in Fig.~\ref{fig:ffMD-Lam-DH}, we show the corresponding results for dark-halo configurations, where the mass-scaled $f$-mode frequencies associated with the dark matter component are plotted as a function of $\Lambda_t$, along with their deviations from the same empirical relation. From both figures, it is evident that the mass-scaled $f$-mode frequencies in the presence of dark matter exhibit significant deviations from the empirical relation established for standard neutron stars without dark matter. This systematic deviation---especially when observed across a wide range of interaction strengths and DM mass fractions---could serve as a potential signature of dark matter inside neutron stars.

\begin{figure}[tbp]
\begin{center}
\includegraphics[scale=0.58]{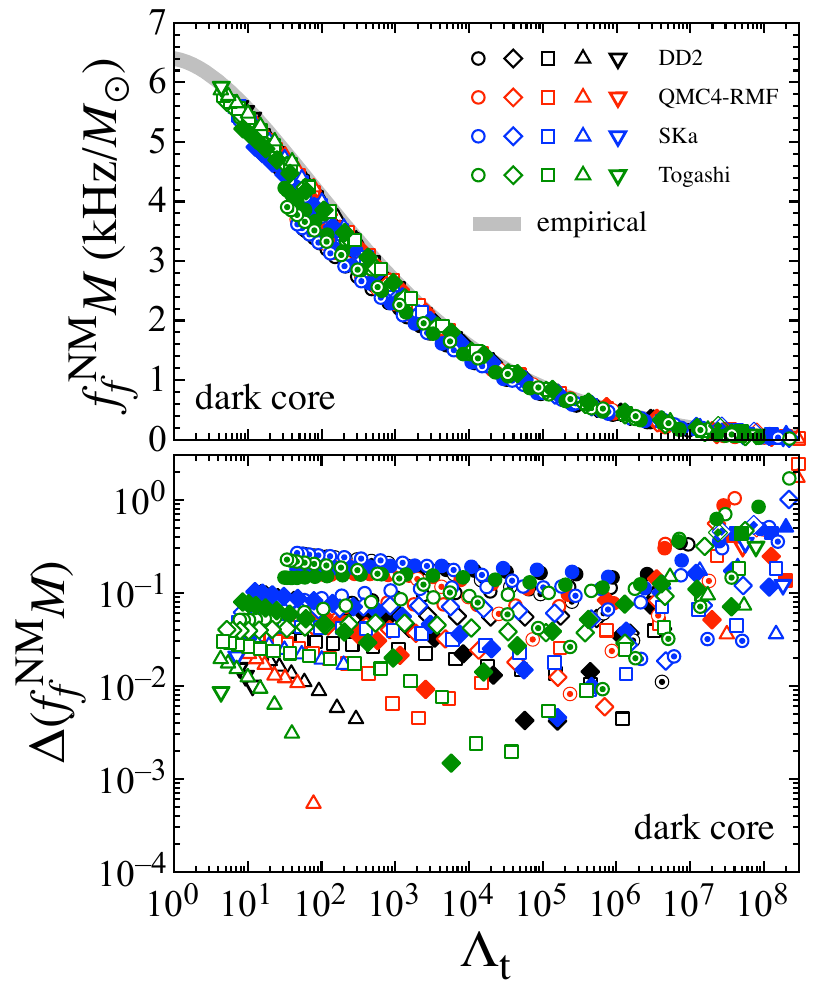} 
\end{center}
\caption{
Mass-scaled $f$-mode frequencies associated with normal matter ($f_f^{\rm NM} M$) are plotted as a function of the dimensionless tidal deformability $\Lambda_t$ for dark core configurations constructed with various nuclear EOSs and dark matter parameters. Circles, diamonds, squares, triangles, and inverted triangles correspond to $g_\chi/m_v = 0.02$, 0.04, 0.06, 0.08, and 0.10 MeV$^{-1}$, respectively, while open, filled, and double symbols indicate dark matter mass fractions of $M_{\rm DM}/M=0.2$, 0.4, and 0.6. The thick solid line shows the empirical relation of mass-scaled $f$-mode frequencies with the dimensionless tidal deformability for neutron star models without dark matter. The bottom panel displays the relative deviation of $f_f^{\rm NM} M$ from this empirical relation. 
}
\label{fig:ffMD-Lam-DC}
\end{figure}

\begin{figure}[tbp]
\begin{center}
\includegraphics[scale=0.58]{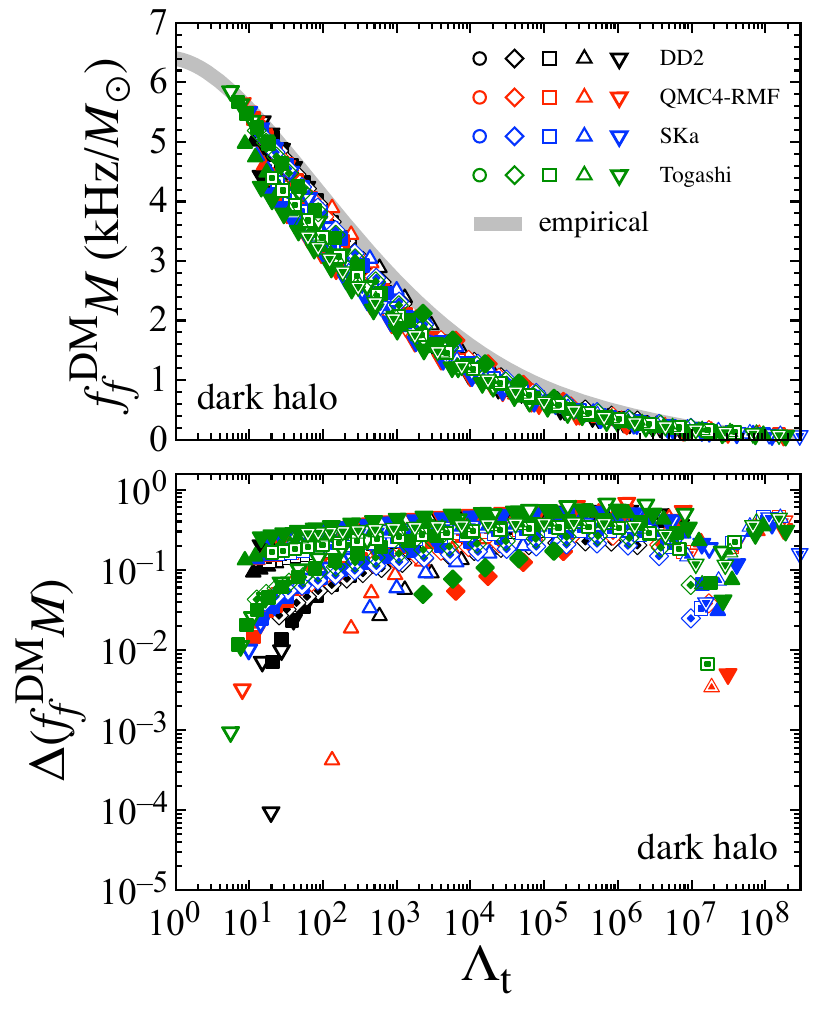} 
\end{center}
\caption{
Same as Fig.~\ref{fig:ffMD-Lam-DC}, but focusing on the mass-scaled $f$-mode frequencies associated with dark matter ($f_f^{\rm DM} M$) for the dark halo configurations.
}
\label{fig:ffMD-Lam-DH}
\end{figure}

\section{Conclusion}
\label{sec:Conclusion}

While the presence of dark matter inside neutron stars remains observationally uncertain, it is plausible that dark matter could accumulate in their interiors due to its dominant gravitational interaction with normal matter. If dark matter is indeed present, it can significantly alter the internal structure and oscillation frequencies of neutron stars, depending on the properties of the dark matter model and the EOS of normal matter. To probe potential signatures of dark matter in neutron stars, we have systematically investigated whether well-established universal relations---originally derived for ordinary neutron stars---remain valid in the presence of dark matter. Specifically, we examined the relations between (i) the dimensionless tidal deformability, $\Lambda_t$, and compactness, $M/R$, (ii) the mass-scaled $f$-mode frequency, $f_fM$, and compactness, and (iii) $f_fM$ and $\Lambda_t$, focusing on a self-interacting fermionic dark matter scenario within the two-fluid formalism. As a result, we find that the relation between $\Lambda_t$ and $M/R$ and the relation between $f_fM$ and $\Lambda_t$ exhibit significant deviation from the universal relations established for neutron star models without dark matter. These deviations could serve as potential observational signatures for inferring the presence of dark matter inside neutron stars. On the other hand, we also find that the mass-scaled $f$-mode frequencies associated with normal (dark) matter excited in the dark core (dark halo) configurations can still be well expressed with the universal relation originally derived for the neutron stars without dark matter. However, we could not identify any robust universal relation for the $f$-mode frequencies associated with dark (normal) matter in the dark core (dark halo) configurations, which would otherwise be valuable for extracting further information about the interior composition. Finally, while this study relied on the Cowling approximation to compute gravitational wave frequencies, we plan to extend our analysis to include full metric perturbations, enabling a more quantitative and accurate investigation of the asteroseismology of dark matter admixed neutron stars.

\begin{acknowledgements}
This work is supported in part by Japan Society for the Promotion of Science (JSPS) KAKENHI Grant Numbers 
JP23K20848         
and JP24KF0090. 
\end{acknowledgements}



\appendix
\section{Universal relations involving $\Lambda_t$ for neutron stars without dark matter}   
\label{sec:appendix_1}

In this appendix, we derive the universal relations involving the dimensionless tidal deformability, $\Lambda_t$, for the neutron star models without dark matter---specifically, ${\cal C}-\Lambda_t$ and $\Lambda_t-f_fM$ relations, where  ${\cal C}$ is the stellar compactness, and $f_fM$ denotes the mass-scaled $f$-mode frequency. In particular, to investigate whether the $f$-mode frequencies computed using the Cowling approximation depend on the presence of dark matter, we explicitly derive the $\Lambda_t-f_fM$ relation in the absence of dark matter, adopting the same Cowling approximation. For consistency, we adopt the same EOS set as used in the Appendix of Ref.~\cite{SK25}. The ${\cal C}-\Lambda_t$ relation was previously presented in Ref.~\cite{SK21}, but the QMC-RMF4 EOS was not included, and the fitting formula was not explicitly provided. As shown in the top panel of Fig.~\ref{fig:C-L}, the relation between the dimensionless tidal deformability and stellar compactness exhibits only a weak dependence on the choice of EOS. Using the dimensionless tidal deformability for several stellar models constructed with various EOS, we derive the fitting formula presented in Eq.~(\ref{eq:C_Lam}). As shown in the bottom panel of Fig.~\ref{fig:C-L}, this fitting formula can estimate the dimensionless tidal deformability for the canonical neutron stars ($\Lambda_t\lesssim 1000$) without dark matter to within $\sim 20\%$ accuracy, provided the stellar compactness is known. 

In a similar way, it is known that the mass-scaled $f$-mode frequency, $f_fM$, can be expressed as a function of the dimensionless tidal deformability, $\Lambda_t$, with minimal dependence on the EOS---at least in the absence of dark matter, e.g.,~\cite{SK21}. As shown in the top panel of Fig.~\ref{fig:L-ffM}, this universal relation holds even when the $f$-mode frequencies are computed using the Cowling approximation, allowing us to derive the fitting formula given in Eq.(\ref{eq:ffM_Lam}). As demonstrated in the bottom panel of Fig.~\ref{fig:L-ffM}, this formula can predict the value of $f_fM$ for canonical neutron star models (i.e., $\Lambda_t \lesssim 1000$) without dark matter to within a few percent accuracy.

\begin{figure}[tbp]
\begin{center}
\includegraphics[scale=0.58]{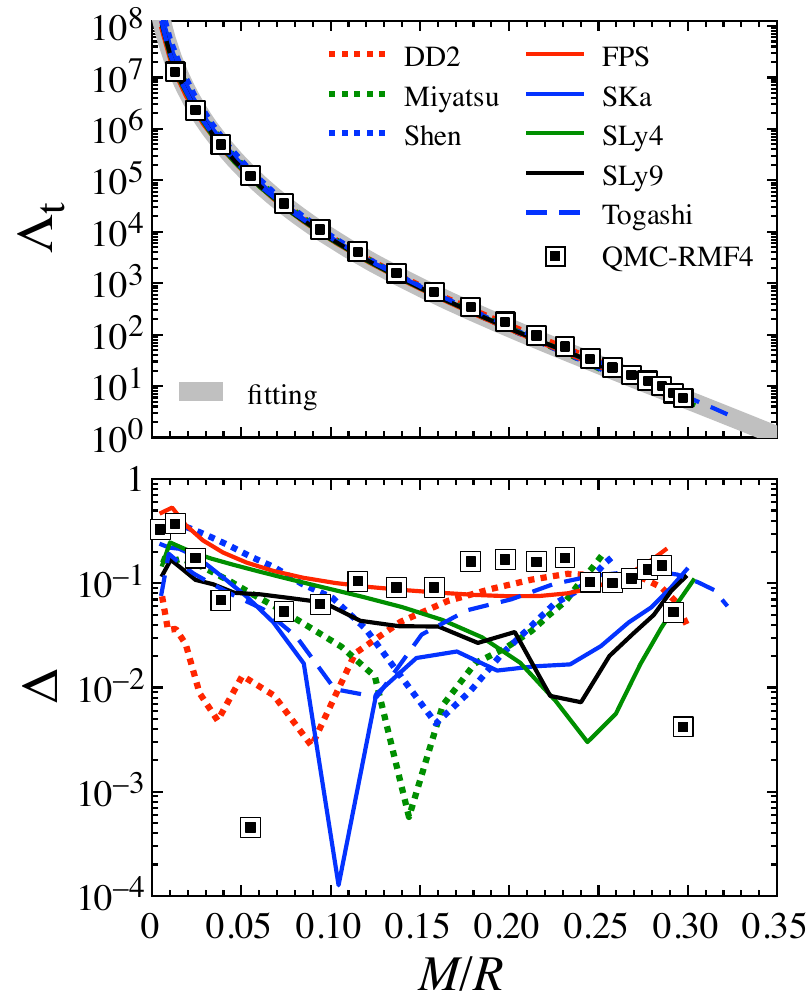} 
\end{center}
\caption{
In the top panel, the dimensionless tidal deformability, $\Lambda_t$, is shown as a function of the stellar compactness, ${\cal C}=M/R$, for various neutron star models without dark matter. The EOS adopted here is the same as in the Appendix of Ref.~\cite{SK25}. The thick-solid line represents the empirical ${\cal C}-\Lambda_t$ relation with the fitting formula given by Eq.~(\ref{eq:C_Lam}). The bottom panel displays the relative deviation of each model’s $\Lambda_t$ from the fitted empirical curve. 
}
\label{fig:C-L}
\end{figure}

\begin{figure}[tbp]
\begin{center}
\includegraphics[scale=0.58]{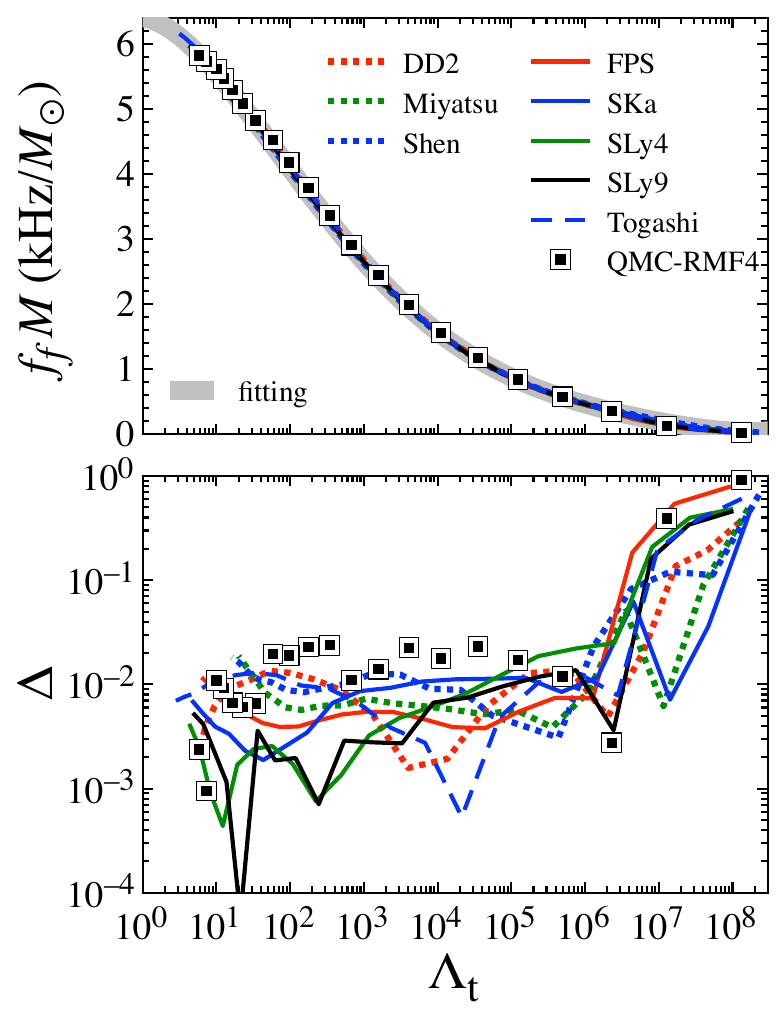} 
\end{center}
\caption{
The top panel shows the relation between the mass-scaled $f$-mode frequencies, $f_fM$, and the dimensionless tidal deformability for various stellar models without dark matter, where the thick solid line denotes the fitting formula given by Eq.~(\ref{eq:ffM_Lam}). The bottom panel displays the relative deviation of the value of $f_fM$ estimated with the fitting formula from those with the $f$-mode frequencies determined using the Cowling approximation. 
}
\label{fig:L-ffM}
\end{figure}



\end{document}